\documentclass[preprint,aps,floats,nofootinbib,amssymb,superscriptaddress]{revtex4}
\pdfoutput=1
\usepackage[sort&compress]{natbib}
\usepackage{epsf,epsfig}
\usepackage{amsmath}
\usepackage{hyperref}
\usepackage{subfigure}
\usepackage{graphicx}
\usepackage{color}
\usepackage{mathrsfs}

\newcommand{\be}{\begin{equation}}
\newcommand{\ee}{\end{equation}}
\newcommand{\ba}{\begin{array}}
\newcommand{\eaq}{\end{array}}
\newcommand{\bea}{\begin{eqnarray}}
\newcommand{\eea}{\end{eqnarray}}
\newcommand{\no}{\nonumber}
\newcommand{\nn}{\nonumber}

\newcommand{\bi}{\begin{itemize}}
\newcommand{\ei}{\end{itemize}}
\newcommand{\bal}{\begin{aligned}}
\newcommand{\eal}{\end{aligned}}

\newcommand{\fracp}[2]{ \left( \frac{#1}{#2} \right)}
\newcommand{\lrp}[1]{ \left( #1 \right) }

\makeatletter

\begin{document}


\title{\vspace*{1.in} Interpreting a 2 TeV resonance in WW scattering}

\author{Pere Arnan}\affiliation{Departament d'Estructura i Constituents de la Mat\`eria}
\author{Dom\`enec Espriu}\affiliation{Departament d'Estructura i Constituents de la Mat\`eria}
\author{Federico Mescia}\affiliation{Departament de F\'\i sica Fonamental\\Institut de Ci\`encies del Cosmos (ICCUB), \\
Universitat de Barcelona, Mart\'i Franqu\`es 1, 08028 Barcelona, Spain}

\vspace*{2cm}

\thispagestyle{empty}

\begin{abstract}
A diboson excess has been observed ---albeit with very limited statistical significance--- in 
$WW$, $WZ$ and $ZZ$ final states at the LHC experiments using the accumulated 8 TeV data. 
Assuming that these signals are due to resonances resulting from an extended symmetry 
breaking sector in the standard model and exact custodial symmetry
we determine using unitarization methods the values of the relevant low-energy constants 
in the corresponding effective Lagrangian. Unitarity arguments
also predict the widths of these resonances. We introduce unitarized form factors 
to allow for a proper treatment of the resonances in Monte Carlo generators and
a more precise comparison with experiment.

\end{abstract}

\maketitle

\section{Introduction}
In a series of recent papers~\cite{bcn1,bcn2,bcn3,mad1,mad2,mad3} the relation between 
the coefficients of an effective Lagrangian 
parameterizing an extended electroweak symmetry breaking sector (EWSBS) and the appearance of 
narrow resonances in several isospin and angular momentum channels involving the scattering
of longitudinally polarized $W, Z$ bosons has been clearly established.
It was found that, except for a small set of points in the space of parameters very close to the 
minimal Standard Model (MSM) values, resonances with these characteristics should appear. 
In fact it was argued that 
detecting such resonances, if ever found, could provide an indirect but effective way of determining 
anomalous triple and quartic gauge boson vertices.

The connection between resonances and 
coefficients of the effective EWSBS Lagrangian is not based on a fully rigorous mathematical theorem,
but it is amply supported by a wealth of experience on strong interactions and unitarization techniques
in effective theories~\cite{unit}. In the present context results  have been provided 
by two different groups.
In ~\cite{bcn1,bcn3} some of the present authors found by using the inverse amplitude method (IAM) of unitarization
the relation between the characteristics of the 
first resonance in the various $IJ$ channels ($I=$ custodial isospin) and the value 
of the coefficients of the effective Lagrangian.
The analysis was done making only as minimal as possible an usage of the equivalence theorem~\cite{ET,esma}
as this is known to be prone to substantial corrections at low values of $s$. 
The Madrid group~\cite{mad1,mad2,mad3}  making use of the equivalence theorem have
also been able to determine the connection between resonances and departures from the MSM at an
effective Lagrangian level. The agreement between the two independent set of calculations
is excellent whenever they can be compared. In addition the Madrid group has done a careful analysis 
of different unitarization methods \cite{mad3}. 

Unitarization leads to various resonances
depending on the values of the effective couplings. In addition there is an ample region of parameter space 
ruled out as viable effective theories, something that is not a
surprise to effective theory practitioners~\cite{excluded}. While there is certainly some room for some 
quantitative differences between different unitarization methods, 
the results are generally believed to be fairly accurate.

In the present discussion by unitarization we refer to the reconstruction of a unitary amplitude using 
tree-level plus one-loop results. Several works considering the so-called 
tree level unitarity (i.e. the requirement
that amplitudes of the kind considered here do not grow with $s$) already exist \cite{unitar}. 

Recently the experimental collaborations ATLAS and CMS have reported~\cite{atlas,cms} a modest 
excess of diboson events peaking around the 2 TeV region. ATLAS looks for the invariant mass distribution of a pair of jets
that are compatible with a highly boosted $W$ or $Z$ boson. CMS combines dijet and final states with one or two 
leptons and concludes that there is a small excess around 1.8 TeV but with less statistical significance.
In what follows we shall use the ATLAS resultss assuming a mass for a putative resonance in the
range 1.8 TeV $< M <$ 2.2 TeV.

In hadronic decays such as the ones used by ATLAS it is not always possible
to establish the nature of the jet ($W$ or $Z$)~\cite{expdiscussion}. 
Yet the experimental collaboration feel confident
enough to claim that the signal is apparently present 
in the three channels $WW$, $WZ$ and $ZZ$. Assuming exact custodial symmetry this would
suggests that the resonance could not have $I=0$ as this would not contribute in the $s-$channel to 
$WZ$ scattering, where the signal appears to be stronger.

However, elementary isospin arguments forbid a resonant 
contribution with $I=1$ in processes with a $ZZ$ final state. Therefore
assuming exact custodial symmetry, whether
the resonance has either $I=0$ or $I=1$ one of the `observed' channels
must have necessarily been misidentified~\cite{expdiscussion}. The alternative to accepting 
$O(1)$ custodial breaking would be to contemplate a 
resonant $I=2$ state (contributes to all final
states), but we regard this as unlikely for the reasons described in detail in \cite{bcn3,falk}
 (but see \cite{GM} where an elementary $I=2$ state is introduced).

In this letter we shall contemplate the two hypothesis $I=0, J=0$ and $I=1,J=1$ and 
use the IAM to derive a very restrictive bound on a combination of two coefficients of the effective 
Lagrangian. 
In addition we will be able to approximately determine the widths of these putative resonances. 
The allowed regions in parameter space partly overlap; namely there are regions with
{\em both} a scalar and vector resonances (this would of course help to
explain the excess in {\em all} channels). We will comment on the respective
possible widths and masses. We will see that the range of masses contemplated here would lead to
a severe reduction in the range of variation of the low-energy constants
providing precious information to disentangle the class of underlying physics that one
could be contemplating.

One salient characteristic of the resonances found in the mentioned unitarization analysis is that
they are {\em very narrow}, something that runs contrary to the intuition of many practitioners 
in strongly interacting theories. This comes about because of the strong but partial unitarization
that a Higgs at $M_H=125$ GeV brings about. By construction these resonances couple {\em only} 
to $W$ and $Z$ bosons. Together with the assumption of exact custodial symmetry, 
this is the only hypothesis in our analysis.

\section{Constraining the effective Lagrangian coefficients}
The effective Lagrangian whose unitarized amplitudes we will consider is
\bea
\label{eq:1}
\mathcal{L} & = &  - \frac{1}{2} {\rm Tr} W_{\mu\nu} W^{\mu\nu} - \frac{1}{4} {\rm Tr} B_{\mu\nu} B^{\mu\nu} 
+ \frac{1}{2} \partial_{\mu} h \partial^{\mu} h  - 
\frac{M_H^2}{2} h^{2} - d_{3} (\lambda v)  h^{3} -  d_{4} \dfrac{\lambda}{4} h^{4} \\ \nn 
& & + \frac{v^{2}}{4} \lrp{1+2 a\fracp{h}{v}+ b \fracp{h}{v}^{2}+...} {\rm Tr}\, D_{\mu}U^{\dagger}D^{\mu}U 
+ \sum a_{i} \mathcal{O}_i\,.
\eea
where 
\be
\label{eq:2}
U = \exp \lrp{i~\frac{w \cdot \tau}{v} }\quad\text{and,}\quad D_{\mu} U =  \partial_{\mu}U + 
\frac{1}{2} i g W_{\mu}^{i} \tau^{i} U - \frac{1}{2} i g' B_{\mu}^{i} U \tau^{3}.
\ee
The $w$ are the three Goldstone of the global group $SU(2)_L\times SU(2)_R\to SU(2)_V$. This symmetry breaking 
is the minimal pattern to  provide the longitudinal components to the $W^\pm$ and $Z$ and emerging from phenomenology. 
The Higgs field $h$ is a gauge and $SU(2)_L \times SU(2)_R$ singlet and the $\mathcal{O}_i$ are a set of higher
dimensional operators. In an energy expansion and at the next-to-leading order it is sufficient to
consider the $O(p^4)$ operators. 
This formulation is strictly equivalent to 
others where the Higgs is introduced as part of a complex doublet, as $S$-matrix elements are independent
of the parameterization.  
 
The operators $\mathcal{O}_{i}$ include the complete set of operators defined e.g. 
in \cite{bcn1,ECHL,dob}. We will be interested in $WW$ scattering and 
work in the strict custodial limit. Therefore, only a restrict number of operators have to be considered;
namely of the possible 13 $O(p^4)$ operators only two $O_4$
and $O_5$ will contribute to $W_LW_L$ scattering\footnote{It should be obvious that when we talk
about $WW$ or $W_LW_L$ scattering we refer generically to any scattering of vector bosons. 
Concrete processes are specified when needed.}
 in the custodial limit:
\be
\mathcal{O}_{4} = {\rm Tr}\left[ V_{\mu}V_{\nu} \right]{\rm Tr}\left[ V^{\mu}V^{\nu} \right] 
\qquad
\mathcal{O}_{5} = {\rm Tr}\left[ V_{\mu}V^{\mu} \right]{\rm Tr}\left[ V_{\nu}V^{\nu} \right],
\ee
where $V_{\mu} = \left( D_{\mu} U \right) U^{\dagger}$.
We could easily extend the analysis to include non-custodial contributions, but we see little or no reason
to do so at present.

The parameters $a$ and $b$ control the coupling of the Higgs to the gauge sector~\cite{composite}. 
Couplings containing higher powers of $h/v$ do not enter $WW$ scattering and they have not
been included in (\ref{eq:1}). The two additional parameters
 $d_{3}$, and $d_{4}$ parameterize the three- 
and four-point interactions of the Higgs field\footnote{This is not the most 
general form of the Higgs potential and in fact additional counter-terms are needed beyond the 
Standard Model\cite{mad1}, but this does not affect the subsequent discussion for $W_LW_L$ scattering}. 
The MSM  case corresponds to setting $a=b=d_{3}=d_{4}=1$ in Eq. (\ref{eq:1}). 
Current LHC results give the following bounds for $a$, $a_{4,5}$:
\be
\label{eq:bound_a_a4_a5}
a= [0.67,1.33],  \qquad  a_4= [-0.094,0.10], \qquad a_5= [-0.23,0.26]\qquad 
90\%\text{CL}
\ee 
see \cite{Falkowski:2013dza,Concha}
.
Present data clearly favours values of $a$ close to the MSM value ($a=1$). We shall
consider here only this case leaving the consideration other values of $a$ to a forthcoming
publication\footnote{It should be mentioned at this point that considering $a<1$ leaves the 
vector cross-section almost unchanged (although the range of $a_4$ $a_5$ is somewhat modified)
but does increase noticeably the scalar cross-section.}. The parameter $b$ is almost totally
undetermined at present and actually does not play a very relevant role in 
the present discussion. We will assume $b=a^2$ without further adue.

Determining the range of parameters $a_4$ and $a_4$ allowed by assuming a scalar and/or
vector resonance in the range
1.8 TeV $<M<$ 2.2 TeV is the main purpose of the present analysis. It should be mentioned that these
two low-energy constants do not affect at all oblique corrections (quite constrained, see e.g.
\cite{oblique}) nor the triple gauge 
boson coupling:  $a_1$, $a_2$ and $a_3$ are the relevant couplings in the custodial limit to
consider in these contexts. The effective EWSBS Lagrangian nicely disentangles the two kind
of constraints.

We shall not provide here the technical details of the unitarization method we use as they 
have been described in detail elsewhere~\cite{bcn1,bcn3}. 
  
After requiring a resonance in the vector channel with a mass in the quoted range one gets in 
a $a_4-a_5$ plane the region shown on the left in Figure 1 for $a=1$.
An analogous procedure but assuming that the resonance 
is the $I=0,J=0$ channel results in the allowed region in the $a_4-a_5$ plane depicted in
Figure 2.
\begin{figure}[ht!]
\centering
\subfigure[(a)]{\includegraphics[clip,width=0.4\textwidth]{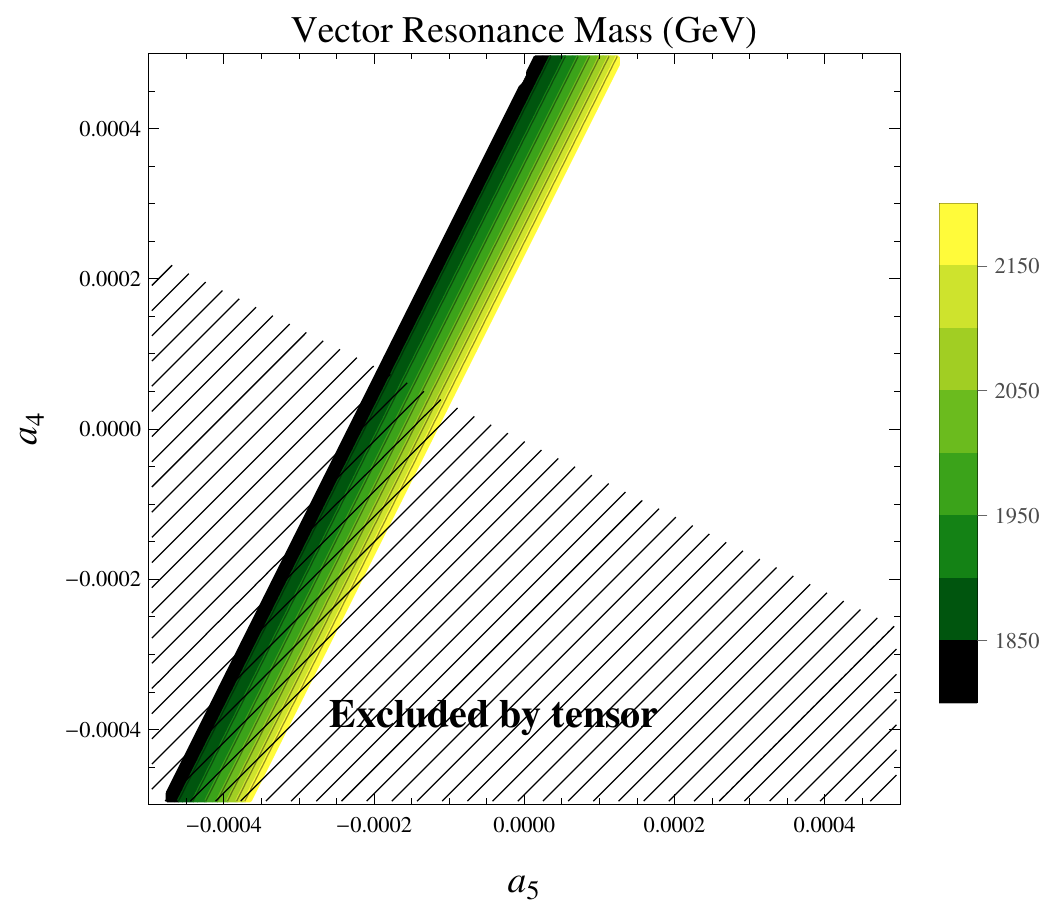}} 
\subfigure[(b)]{\includegraphics[clip,width=0.4\textwidth]{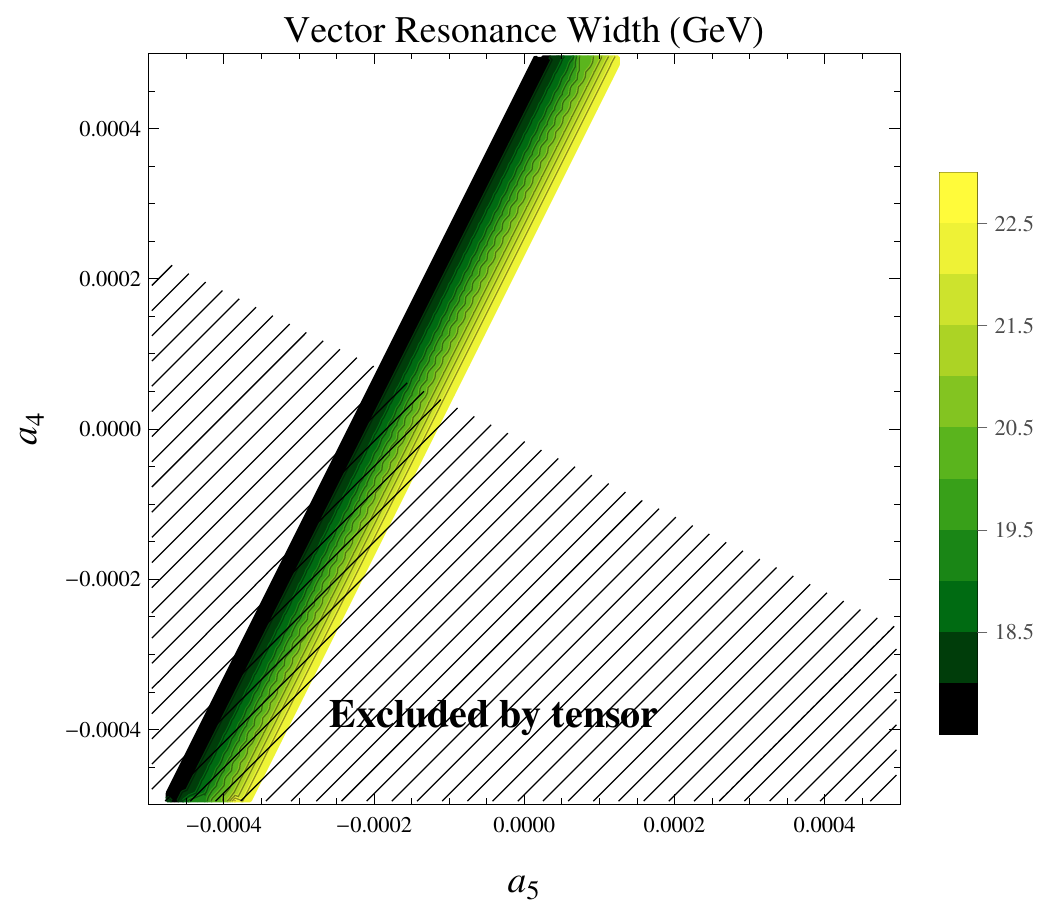}} 
\caption{For $a= 1$ and $b=1$:  (a) allowed values for $a_4$, $a_5$ corresponding to a 
vector resonance
with a mass between 1.8 TeV and 2.2 TeV. Note the extremely limited range of variation that
is allowed in the figure for the low-energy constants. (b) The corresponding widths as predicted
by unitarity using the IAM method. The characteristic value is 20 GeV --- quite narrow for such
a large mass. The dashed area is excluded on causality grounds stemming from the $I=2$ channel.}
\end{figure}
\begin{figure}[ht!]
\centering
\subfigure[(a)]{\includegraphics[clip,width=0.4\textwidth]{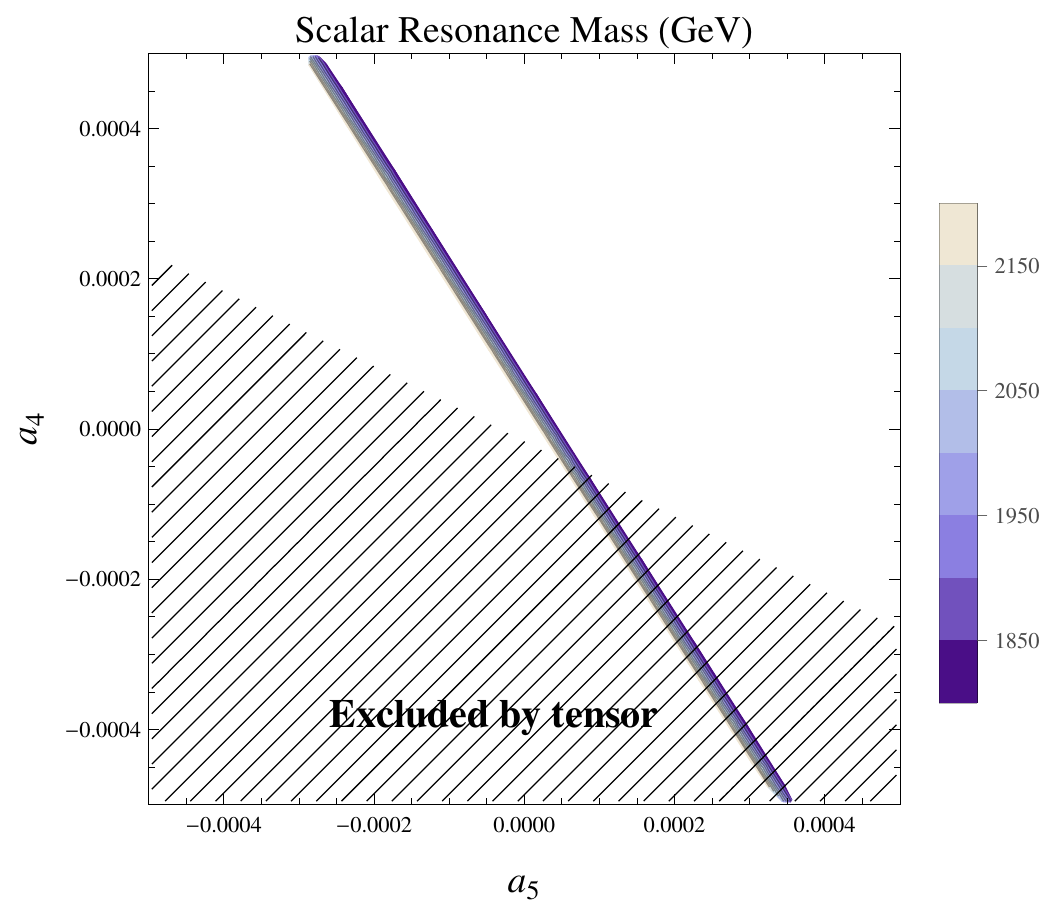}} 
\subfigure[(b)]{\includegraphics[clip,width=0.4\textwidth]{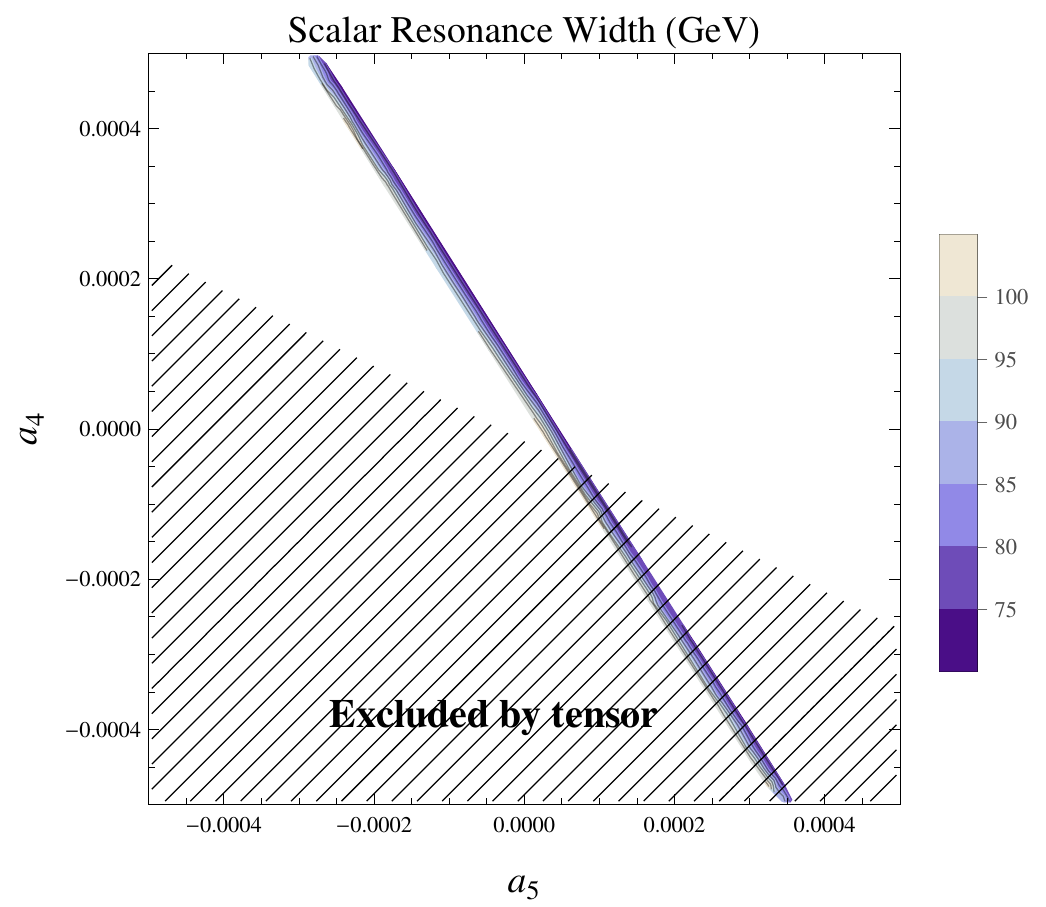}} 
\caption{For $a= 1$ and $b=1$:  (a) allowed values for $a_4$, $a_5$ corresponding 
to a scalar resonance
with a mass between 1.8 TeV and 2.2 TeV. (b) The corresponding widths as predicted
by unitarity using the IAM method; characteristic values are in the 70-100 GeV range.}  
\end{figure}

We would like to emphasize the very limited range of variation for the parameters that is 
shown in Figures 1 and 2. The constants $a_4$ and $a_5$ lay in the small 
region $|a_4|, |a_5| < 5\times 10^{-4}$. (This region includes of course the MSM value
$a_4=a_5=0$, but ---obviously--- there are no resonances there.)

In order to convey a picture of the sort of predictive power of unitarization
techniques we plot in Figure 3
the allowed bands in the broader range $|a_4|,|a_5| < 0.01$ that was considered in a previous
work~\cite{bcn1} as still being phenomenologically acceptable. 
Indeed, setting even a relatively loose bound for the mass of the resonance restricts the range of 
variation of the relevant low-energy constants enormously. In the same Figure 3 we show 
a blown-up of the region where {\em both} a scalar and a vector resonance in this mass range
may coexist. The dashed area is excluded as acceptable for effective EWSBS theories (see \cite{bcn3}). 
\begin{figure}[tb]
\centering
\subfigure[(a)]{\includegraphics[clip,width=0.4\textwidth]{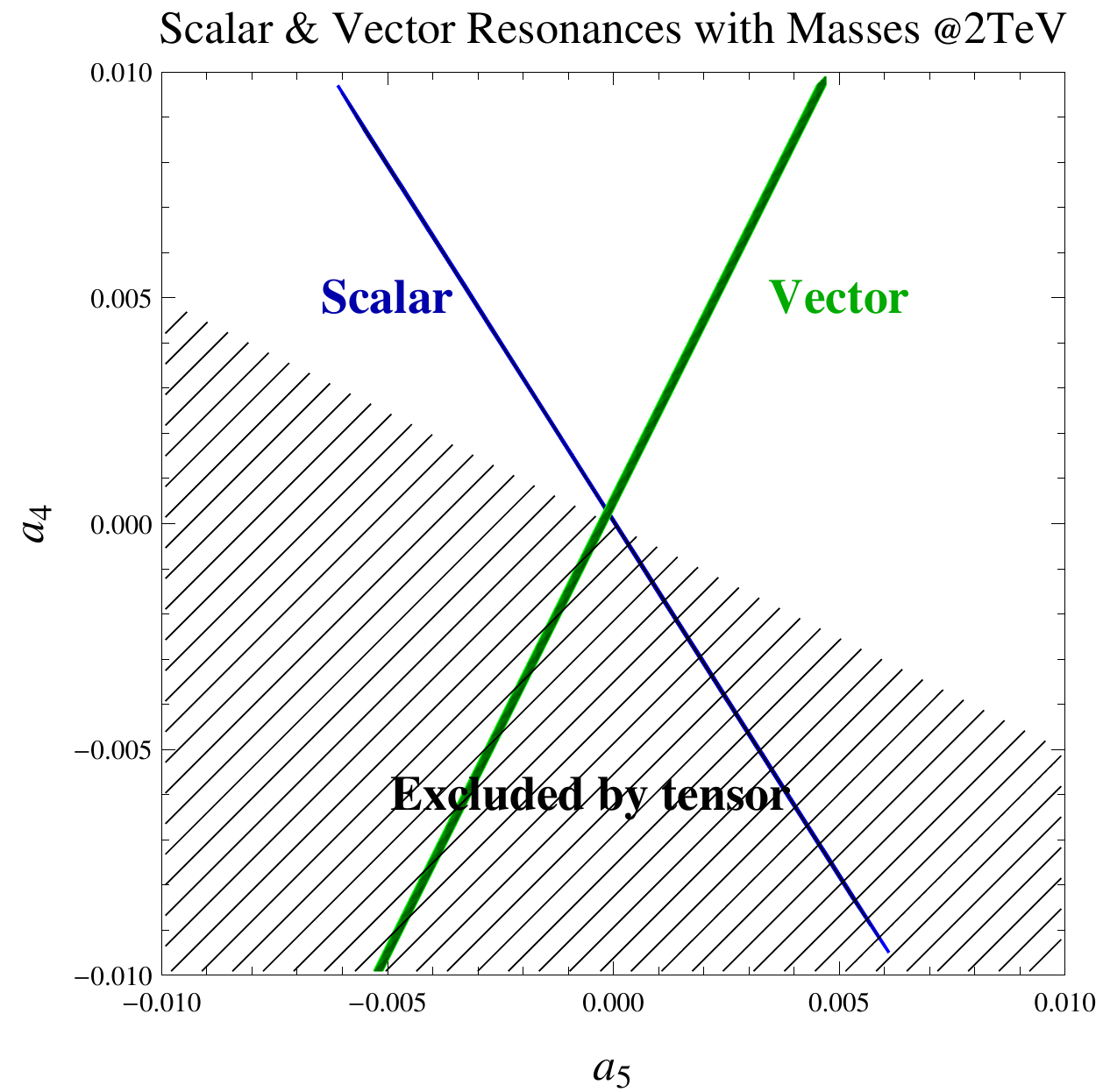}} 
\subfigure[(b)]{\includegraphics[clip,width=0.48\textwidth]{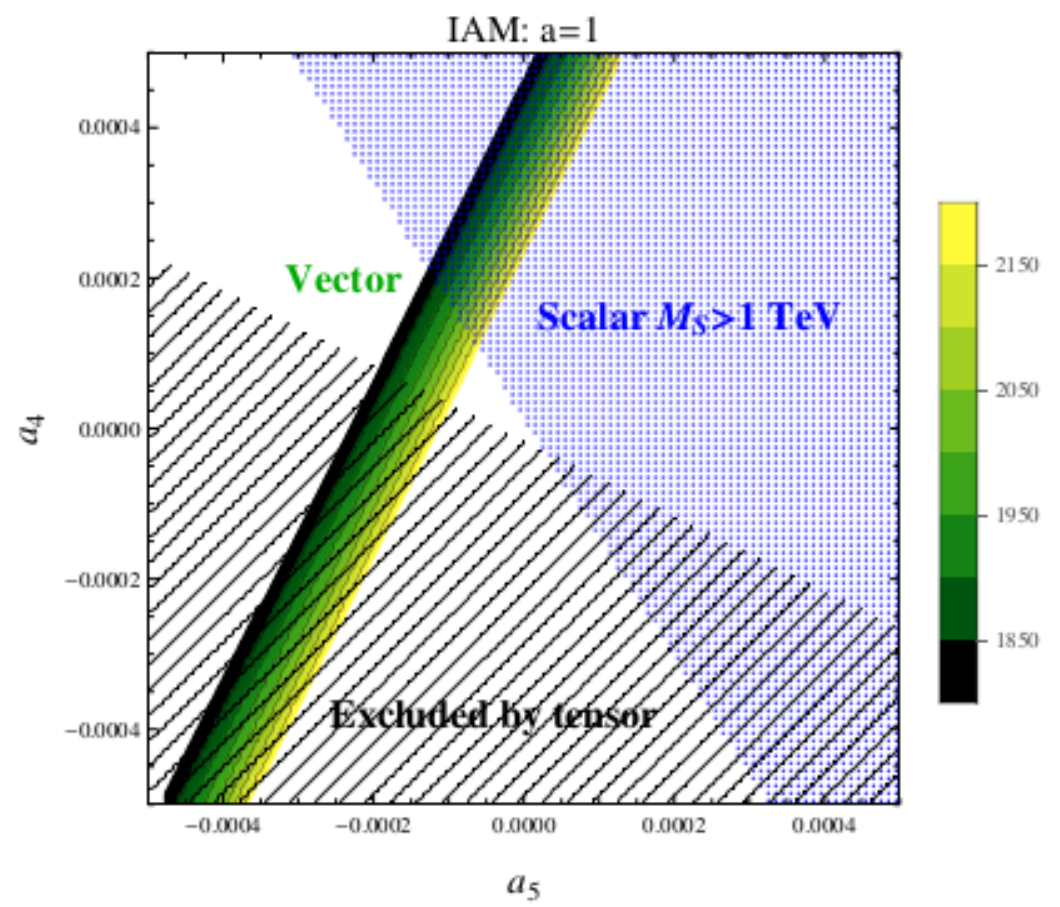}} 
\caption{(a) This plot makes visible how restrictive for the low-energy constants of the 
EWSBS effective Lagrangian becomes the requirement
of yielding a resonance in the 1.8~TeV $<M<$ 2.2~TeV range. The dashed area
is excluded on causality grounds. 
(b) Blow-up of the region of overlap where vector and scalar
resonances may coexist. The broad strip shows
the region of admissible vector resonances with masses in the 1.8-2.2 TeV range. The shaded area
in the upper-right part contains scalar resonances of  mass $>1$ TeV.}  
\end{figure}
\begin{figure}[tb]
\centering
\subfigure[(a)]{\includegraphics[clip,width=0.4\textwidth]{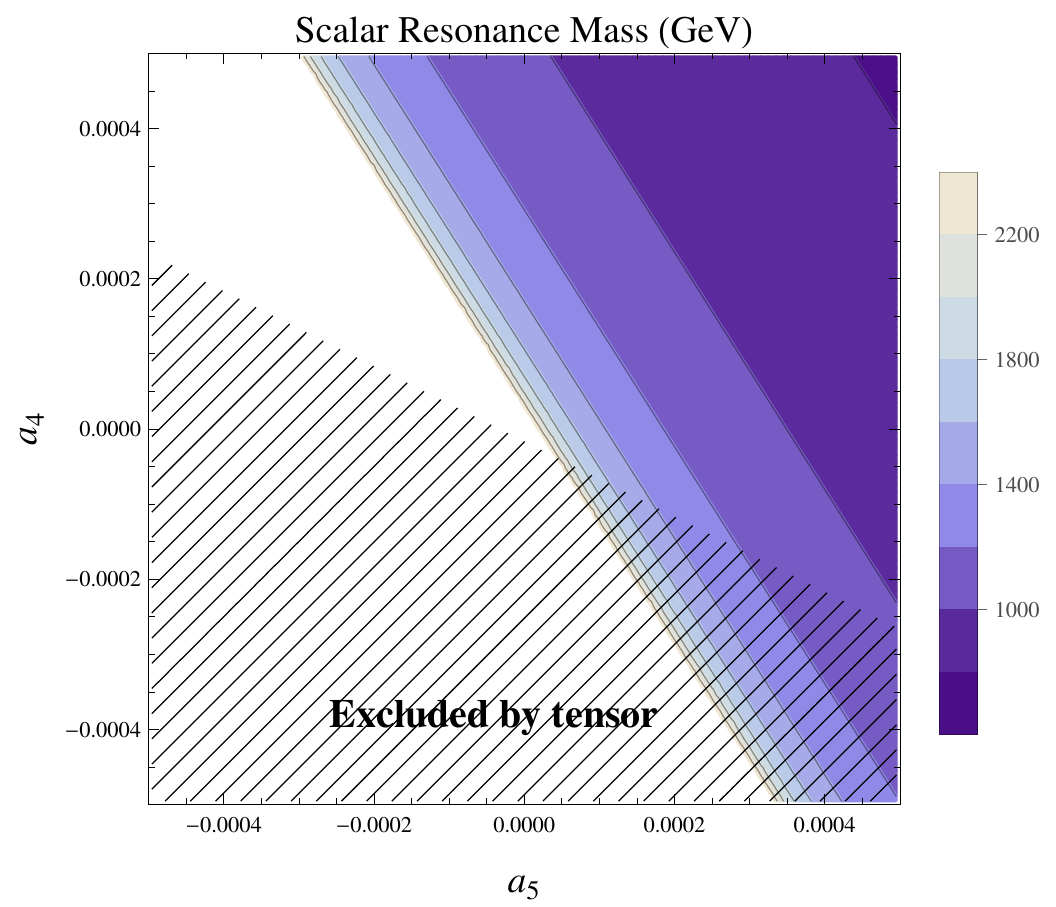}} 
\subfigure[(b)]{\includegraphics[clip,width=0.4\textwidth]{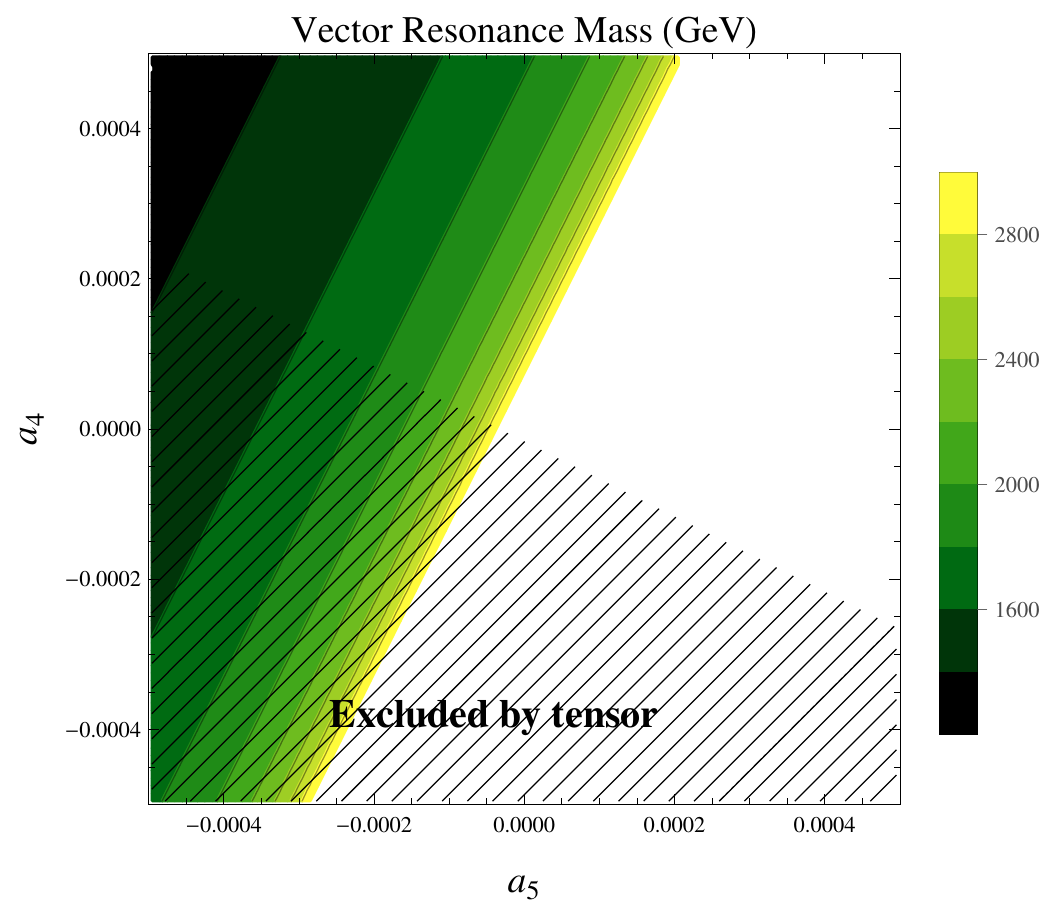}} 
\caption{(a) Viable scalar resonance masses in the region of interest in the
$a_4$-$a_5$ plane for $a=1$ assuming a vector resonance in the 1.8 TeV $<M<$ 2.2 TeV range.
(b) The reverse situation: assuming a scalar mass in the 1.8 TeV $<M<$ 2.2 TeV range and depicting
the possible values for a vector resonance compatible with it.}  
\end{figure}

\section{Experimental visibility of the resonances}
The statistics so far available from the LHC experiments is limited. Searching for new particles in the LHC 
environment is extremely challenging and analyzing the contribution of possible resonances to
an experimental signal is not easy without a well defined theoretical model with definite
predictions for the couplings, form factors, etc.
The IAM method is able not only of predicting resonance masses and widths but
also their couplings to the $W_LW_L$. In \cite{bcn1,bcn3}
the experimental signal of the different resonances was compared to that of a MSM Higgs
with an identical mass. Because the decay modes are similar (in the vector boson channels that is) and 
limits on different 
Higgs masses are very documented this was a rather intuitive way of presenting the cross-section for 
possible EWSBS resonances, but it is not that useful for heavy resonances as the signal of an 
hypothetical Higgs of analogous mass becomes very broad and diluted. 
This point and several others were discussed in detail in \cite{bcn1}. Here we shall give very 
simple estimates of some cross-sections based on the 
Effective W Approximation (EWA) \cite{EWA} in a couple of channels. These estimates 
should be taken as extremely tentative and only
relevant to establish comparisons between different masses and channels. 
In the last section we will introduce form factors and vertex functions
to allow for a proper comparison with experiment. Please note that in the amplitudes
where scalars contribute the contribution of the 125 GeV Higgs is also included.

\begin{figure}[tb]
\centering
\includegraphics[clip,width=0.4\textwidth]{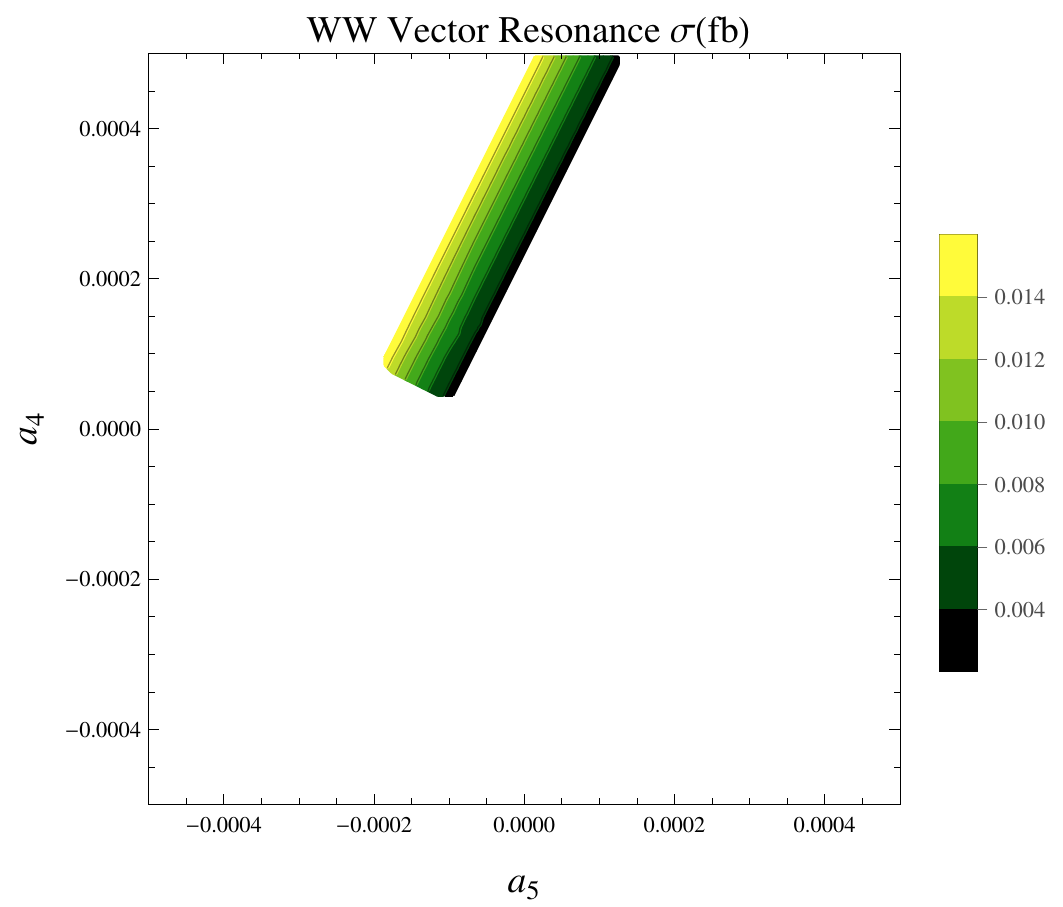} 
\includegraphics[clip,width=0.4\textwidth]{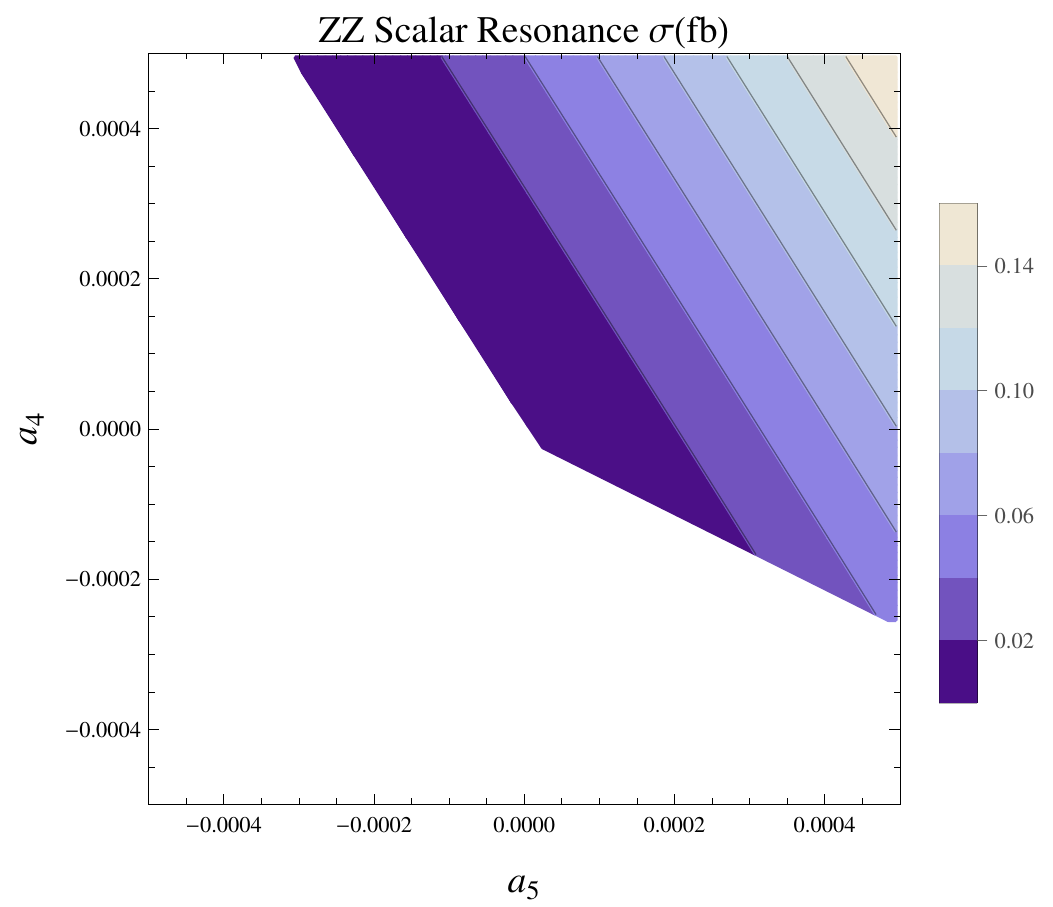}
\caption{Experimental signal of resonances for $a= 1$: 
the resonance cross sections are given in fb,  
the LHC energy has been taken to be 8 TeV and the EWA
approximation is assumed in this calculation. Left: estimated cross section for
the process $W_LW_L\to W_LW_L$ as a function of the parameters $a_4$, $a_5$ due
to a vector resonance. 
Right: cross section for the process $Z_L Z_L\to Z_LZ_L$ due to a scalar resonance. 
The contribution from the 125 GeV Higgs is also included in both cases. \label{signal}}
\end{figure}

Some results for the cross sections are depicted in Figure \ref{signal} for the 
processes $W_L^+W_L^-\to W_L^+W_L^-$ and 
$Z_L Z_L\to Z_LZ_L$.  In the first case we quote the contribution 
from a possible vector resonance only (a scalar resonance is
also possible in this process). In the second case only scalar exchange is possible. 
Note that both diboson production modes
are sub-dominant at the LHC with respect to 
gluon production mediated by a top-quark loop and that the possible resonances in the  
scenario discussed here couple {\em only} to dibosons.

Compared to the preliminary experimental indications, the
results quoted for the cross-sections of these two specific processes are low,  
particulary for vector resonances, but there are several caveats. 
First of all, the EWA tends
to underestimate the cross-sections and it is difficult to assess its validity
in the present kinematical situation. Second, in this region of parameter space the 
cross-sections do change very quickly with only small changes of the parameters thus adding
an element of uncertainty. Finally, the quoted cross sections correspond to considering 
only the interval $s\in [M-2\Gamma, M+2\Gamma]$ so as to have some intuition on the contribution
of the resonance itself. It should also be mentioned that,
as discussed in \cite{bcn1}, there is an 
enhancement in the $W^+ W^-\to W^+W^-$ channel when both the vector and scalar resonances
become nearly degenerate; this is possible in a limited region of parameter space. Also
as previously stated, the scalar channel is enhanced if $a<1$.

Interesting as partial waves for a given process may be, they are not that useful to
implement unitarization in a Monte Carlo generator in order to make detailed
quantitative comparison with experiment. One would need to implement diagrammatic and for 
that one needs vertex functions and propagators wherewith to construct and compute the 
contribution from different topologies. Our proposal to tackle this problem is 
presented next.

\section{Introducing form factors}
The amplitude $A(W^{a}_L(p^{a})+W^{b}_L(p^{b})\to W^{c}(p^{c})_L + W^{d}(p^{d})_L)$ will be denoted
by $A^{abcd}(p^{a},p^{b},p^{c},p^{d})$. Using isospin and Bose symmetries this amplitude can be expressed
in terms of a universal function as 
\be
A^{abcd}(p^{a},p^{b},p^{c},p^{d})= \delta^{ab} \delta^{cd} A(s,t,u) + 
\delta^{ac} \delta^{bd} A(t,s,u)+
\delta^{ad} \delta^{bc} A(u,t,s).
\ee
with $A(s,t,u)= A^{+-00}(p^+,p^-,p^0,p^{\prime 0})$. 
The fixed-isospin amplitudes are given by the following combinations
\bea
\label{eq:fixed_isospin}
T_{0}(s,t,u) & = &  3A(s, t, u) + A(t, s, u) + A(u, t, s) \\ \no
T_{1}(s,t,u) & = &  A(t, s, u) - A(u, t, s) \\ \no
T_{2}(s,t,u) & = &  A(t, s, u) + A(u, t, s) \, . \nn 
\eea
In writing these expressions we assume exact crossing symmetry~\footnote{This remark is
pertinent because amplitudes involving longitudinally polarized bosons are not crossing
symmetric. The formulae can be easily extended to this case but become somewhat more involved
and will not be reported here. See \cite{bcn1}.}.
We also write the reciprocal relations (also assuming exact crossing symmetry)
\bea
\label{eq:fixed_isospin2}
A^{+0+0}(s,t,u) & = & \frac 1 2 T_1(s,t,u)+\frac 1 2 T_2(s,t,u) \\ \no
A^{+-+-}(s,t,u) & = & \frac 1 3 T_0(s,t,u)+\frac 1 2 T_1(s,t,u)+ \frac 1 6 T_2(s,t,u) \\ \no
A^{++++}(s,t,u) & = & T_2(s,t,u)\\ \nn
A^{0000}(s,t,u) & = & \frac{1}{3}T_0(s,t,u)+\frac{2}{3}T_2(s,t,u)\, . \no
\eea
Other amplitudes (such as e.g. $A^{+-00}(s,t,u)$) can be obtained trivially from the previous ones
using obvious symmetries (and crossing symmetry too).

The partial wave amplitudes for fixed isospin $I$ 
and total angular momentum $J$ are defined by
\be\label{eq:legendre}
t_{IJ}(s) = \frac{1}{64\pi} \int_{-1}^{1} d(\cos\theta) P_{J}(\cos\theta) T_{I}(s,t,u) \, ,
\ee
where the $P_{J}(x)$ are the Legendre polynomials and $t = (1 - \cos\theta) (4 M_W^2-s)/2$,
 $u = (1 + \cos\theta) (4 M_W^2-s)/2$ with $M_W$ being the $W,Z$ mass $t_{00}$, $t_{11}$ and
$t_{20}$ are the first non-vanishing partial waves in the present case. The poles in the respective
unitarized partial wave amplitudes dictate the presence or absence of EWSBS 
resonances in the different channels.

\begin{figure}[tb]
\centering
\includegraphics[clip,width=0.9\textwidth]{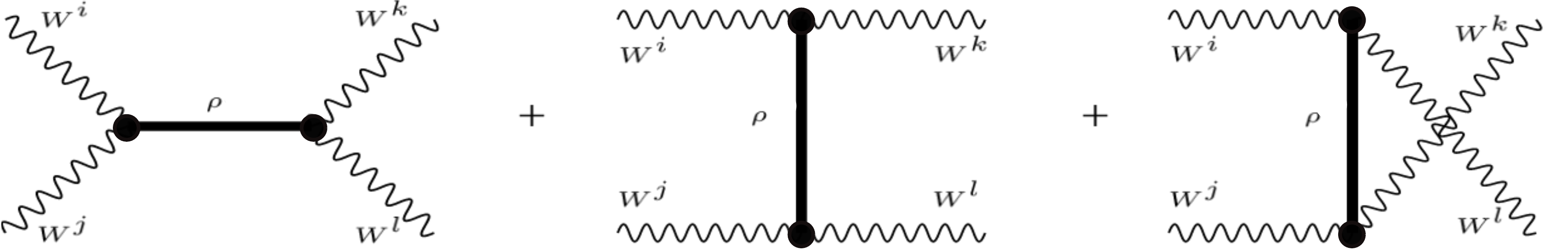} 
\caption{Decomposition of a process in (unitarized) form factors and resonance propagators. \label{diagrams}}
\end{figure}

We would like to express any amplitude as the sum of exchanges of resonances in the $s$, $t$
and $u$ channels, as it is diagrammatically expressed in Figure \ref{diagrams}. That is, we decompose,
say $A^{+0+0}$
\be
A^{+0+0}= \sum_{IJ} (A^{IJ}_s + A^{IJ}_t + A^{IJ}_u )
\ee
Not all $IJ$ receive contributions from all three channels. For example, in the case $A^{+0,+0}$ 
a possible scalar resonance only contributes to the $t$-channel. In addition, not all processes are resonant
in all regions of parameter space, so the above decomposition assumes resonance saturation.
Let us now define the vector form factor as\footnote{CVC has been used.}
\be
\langle W_L^i(p_1)W_L^j(p_2)|J^k_\mu|0\rangle=(p_1-p_2)_\mu F_V(s)\epsilon^{ijk}
\label{genformf}
\ee
where $J^\mu_k$ is the interpolating vector current with isospin index $k$ that
creates the resonance $\rho$ and $F_V(s)$ is the vector form factor.  
From this form factor we derive a vector vertex function $K^\mu $ via the relation
\be
K^\mu(p_1,p_2)=(p_1-p_2)^\mu F_V(s)(s-M_{\text{\tiny{pole}}}^2)
\label{vertexF}
\ee

Let us focus for instance on the amplitude $A^{+0+0}$ that has potentially contributions 
from a vector and a tensor. The IAM does exclude the $I=2$ contribution~\cite{bcn3} so let us
consider $A^{11}_s$ for this process. It can be expressed as
\be
A^{11}_s=K^\mu\hspace{1pt}\frac{g_{\mu\nu}-\frac{k_\mu
k_\nu}{k^2}}{s-M_{\text{\tiny{pole}}}^2}\hspace{1pt}K^{* \nu}=|F_V(s)|^2(s-M_{\text{\tiny{pole}}}^{*2})(-2t-s)=|F_V(s)|^2(s-M_{\text{\tiny{pole}}}^{*2})(-s\cos \theta)
\label{amplitude11}
\ee
where $M_{\text{pole}}=M-i\Gamma/2$. 
Analogous decompositions exist for $A^{11}_t$ and $A^{11}_u$. 
In fact we do not need to consider
$A^{11}_t$ and $A^{11}_u$ at all because assuming exact isospin symmetry $A^{11}(s,t,u)=(-1)^I A^{11}(s,u,t)$. 
Here we assume, and it
is a necessary ingredient of the present approach, that external lines are on-shell. 

On the other hand from unitarization we know that
\be
A^{11}\simeq 96\pi  t_{11}(s)\cos\theta ,
\ee
so neglecting further partial waves it is natural to identify
\be  
|F_V(s)|^2= -\frac{96\pi t_{11}(s)}{s(s-M^{*2}_{\text{\tiny{pole}}})}
\label{ffactor}
\ee
where for $t_{IJ}$ we can use the IAM approximation
\be
t_{IJ} \approx \frac{t_{IJ}^{(0)}}{1-t_{IJ}^{(2)}/t_{IJ}^{(0)}}.
\ee
Although $|F_V|^2$ should of course be real and positive, when using the identification above we
get a tiny imaginary part ($\text{Im}|F_V|^2\sim 10^ {-2} \text{Re}|F_V|^2$) due to the fact 
that we are missing possible channels (including non-resonant contributions) 
and terms in the partial wave expansion. However we can regard
the description of the amplitude via vertex functions and resonance propagators as quite
satisfactory in the regions where resonances are present.

\begin{figure}[tb]
\centering
\includegraphics[clip,width=0.4\textwidth]{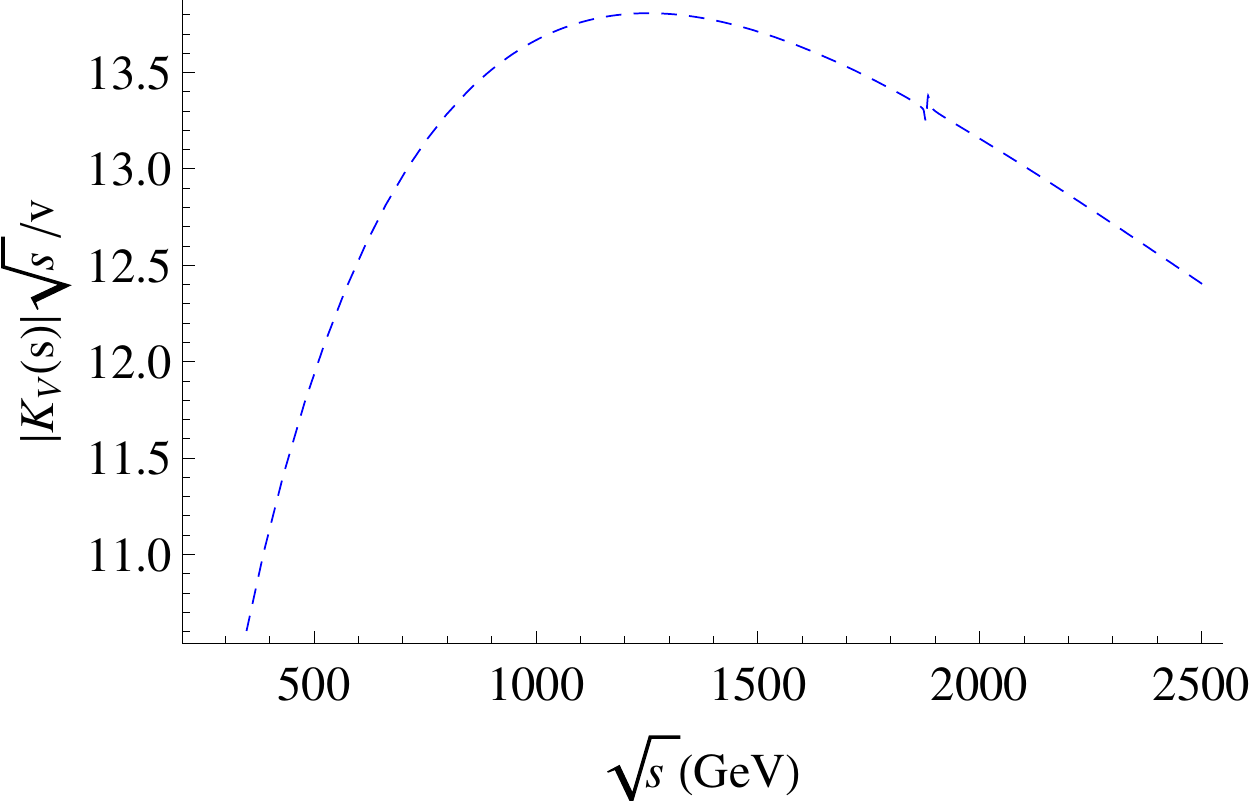}
\includegraphics[clip,width=0.4\textwidth]{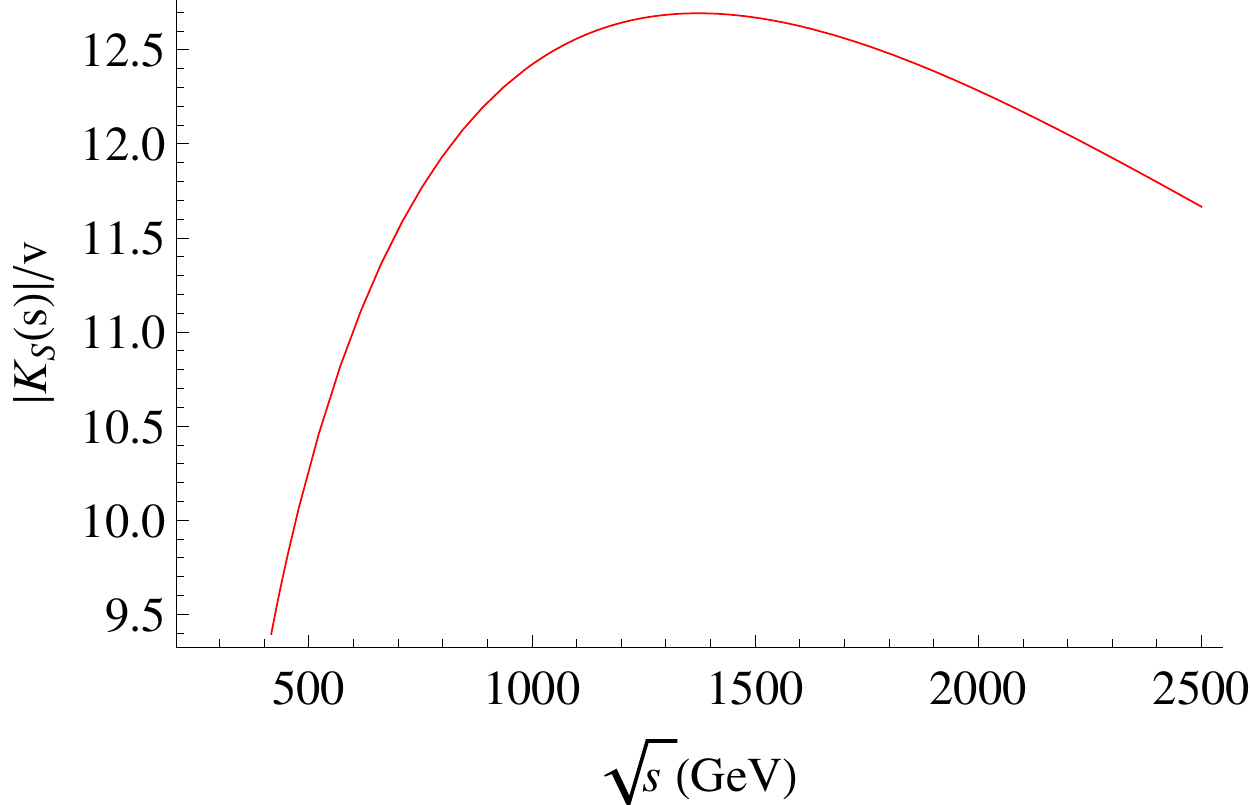} 
\caption{ 
Left: plot of the effective coupling of the vector resonance $K_V(s)\sqrt{s}$ for the
value $M=1881$ GeV corresponding to $a=1$ $a_4=0.0002$,$a_5=-0.0001$ .
Right: plot of the effective coupling for a scalar resonance $K_S(s)$ corresponding to
the same values of $a_4$ and $a_5$ that yields a scalar mass $M=2064$. Note that in both
cases the coupling is quite large, certainly non-perturbative. In fact, on the 
scalar resonance the effective coupling is $\sim 30$ times the coupling of 
a MSM Higgs with identical mass. \label{comparison}}
\end{figure}

Neglecting the gauge boson mass (quite justified at 2 TeV) unitarity requires the form factor 
to obey the following relation within a vector dominance region~\cite{dob}
\be
\text{Im} F_V(s) = t_{11}^*(s) F_V(s).
\label{unitarity}
\ee
Equation (\ref{unitarity}) allows us to extract the phase of $F_V(s)$. Thus, combining the phase 
and the modulus we obtain the vector form factor
\begin{equation}
F_V(s)=|F_V(s)|\exp\left(i \arctan\frac{\text{Re}t_{11}}{1-\text{Im}t_{11}}\right)\,.
\end{equation} 
Similar techniques could allow us to
define a unitarized scalar form factor $F_S(s)$ and a vertex function 
directly derived from the unitarized amplitude that in this channel is
\be
A^{00}\simeq 32\pi  t_{00}(s)
\ee
and assuming resonance dominance.
In Figure \ref{comparison} we plot the vertex functions $K_V(s)$ and $K_S(s)$ obtained by 
the method just described:
\begin{equation}
|K_V(s)|\sim|F_V(s)||s-M_\text{\tiny pole}^2|,\qquad |K_S(s)|\sim|F_S(s)||s-M_\text{\tiny pole}^2|.
\end{equation} 
Note that the function $K_V(s)$ is dimensionless while $K_S(s)$ has units of energy. However
for vector resonances, the effective coupling is typically $K_V(s)\sqrt{s}$ (see the
expression for the form factor and the associated Feynman rule). In the last figure we 
plot these effective couplings normalized to the scale $v$. The contribution to the
form factor from the 125 GeV Higgs is negligible around the scalar resonance at $~ 2$ TeV.

Once we feel confident that the combination of resonant propagators and the vertex functions
just given reproduces very satisfactorily the unitarized amplitudes we can pass on
this information to Monte Carlo generator practitioners to implement
these form factors in their favorite generator. 

The expressions for $M_{\text{\tiny{pole}}}$, $t_{00}(s)$ and $t_{11}(s)$ needed to reproduce the
diagrammatic expansion for the
various values of $a$ and $a_4$, $a_5$ can be found in \cite{bcn1,bcn2,bcn3} (and 
\cite{mad1,mad3} if a full use of the equivalence theorem is made\footnote{Please note
that $t$-channel $W$ exchange is not included in some of these works.}). 
Further details will be provided in a forthcoming extended publication.

\section{Conclusions}
To conclude, we have extracted the values of the low-energy constants $a_4$ and $a_5$ of
the effective Lagrangian describing an extended electroweak symmetry breaking sector
 assuming  (iso)vector dominance and/or (iso)scalar dominance with a 
mass in the range 1.8 TeV $<M<$ 2.2 TeV, as it would be the case if one considers
the preliminary results coming from the LHC experiment to be a hint of the
existence of new $W_LW_L$ interactions. The calculation was performed in the framework of the inverse
amplitude unitarization method. We derived the widths of such resonances, which turn 
out to be quite narrow. We also speculated on the possibility of more than one resonance being
present compatible with the derived bounds on $a_4$ and $a_5$ (something that is favoured by custodial 
symmetry considerations). The given range of masses restrict enormously
the admissible values for $a_4$ and $a_5$ ---surely a consequence of this mass scale being relatively
close to the natural cut-off of the effective theory ($\sim 3$ TeV). The cross-sections obtained using the
Effective W approximation are however two low, particularly for vector resonances, and this may
eventually prove bad news for resonances of the kind considered here. 
However we regard estimates based on the EWA 
as being too preliminary at this point.  

To overcome this difficulty we proposed a diagrammatic method to deal with resonances 
in regions of parameter space
in the effective Lagrangian  where the former are assumed to dominate. We derived the
corresponding form factors and vertex functions. The agreement with the full amplitude is 
very good and we understand that the technique that we introduce here may be
useful to deal with the type of resonances that may emerge in EWSBS. We hope that this 
will trigger interest from our experimental colleagues to incorporate this seemingly consistent
unitary procedure in their generators to allow for a proper theory-experiment comparison. In fact
having a reliable estimate of the resonances cross sections in the region of interest is probably
the most urgent task. 

The apparent signal coming from the LHC experiments has triggered a flurry 
of activity that has mostly concentrated
in proposing specific models ranging from 
introducing resonances \cite{reso} to the obvious possibility of excited
or left-right symmetric $W^\prime, Z^\prime$ states to more exotic models \cite{modelets}.
Our proposal is somewhat different: it is not primarily aimed
at advancing a definite {\em ad hoc} proposal but rather to help understand if the signal is there
in the first place and at trying to elucidate the properties of the resonance (or resonances)
that might be present in an extended electroweak symmetry breaking sector
in $WW$ scattering. 
We regard the restriction on some coefficients of the effective 
Lagrangian provided by unitarity considerations as non-trivial and, if confirmed, would
undoubtedly play a relevant role in constraining the underlying model.

\section*{Acknowledgements}
We thank R. Delgado, A. Dobado, F. Llanes-Estrada and J.R. Pel\'aez for discussions concerning different aspects of
unitarization and effective lagrangians. D.E. thanks the Perimeter Institute where this work was 
initiated for the hospitality extended to him.
We acknowledge the financial support from projects FPA2013-46570, 2014-SGR-104  
and CPAN (Consolider CSD2007-00042).
Funding was also partially provided by the Spanish MINECO under project
MDM-2014-0369 of ICCUB (Unidad de Excelencia `Maria de Maeztu')


\end{document}